\documentclass{elsart}
\usepackage{epsfig}
\begin{document}
\runauthor{Postnikov, Caciuc and Borstel}
\begin{frontmatter}
\title{Structure Optimization and Frozen Phonons in LiNbO$_3$}
\author{A.~V.~Postnikov, V.~Caciuc and G.~Borstel}
\address{
Universit\"at Osnabr\"uck -- Fachbereich Physik,
D-49069 Osnabr\"uck, Germany
}

\begin{abstract}
The equilibrium ground-state structure of LiNbO$_3$ in the paraelectric and
ferroelectric phases is fully optimized in a first-principles
calculation using
the full-potential linearized augmented plane wave method.
The equilibrium volume, $c/a$ ratio and all
(four, in the ferroelectric phase) internal parameters
are found to be in good agreement with the experimental data.
Frozen phonon calculations are performed for TO-$\Gamma$ phonons
corresponding to the $A_1$ and $A_2$ irreducible representations of
the $R3c$ space group in the ferroelectric phase. The comparison
with available experimental frequencies for the $A_1$ modes
is satisfactory (including the $^6$Li isotope effect),
and the displacement patterns are unambiguously attributed.
For the (Raman inactive) $A_2$ modes,
phonon frequencies and eigenvectors are predicted.
\end{abstract}
\end{frontmatter}

\section*{Introduction}
Due to its various applications in non-linear optics and electro-optics,
the ferroelectric material LiNbO$_3$ is being extensively studied
over decades. Its ferroelectric transition temperature of 1480 K
is among the highest known to date.
The mechanism of the structural phase transition from paraelectric
to ferroelectric phase is still an open question. Temperature dependence
measurements of Raman scattering \cite{johnston} and
far infrared reflectivity in LiNbO$_3$ \cite{servion}
suggest the displacive type phase transition. In contradiction with
this picture, the absence of the $A_1$ mode softening reported
in certain papers \cite{o-d}
is an indication towards an order-disorder type phase transition.

\begin{figure}
\centerline{\epsfig{file=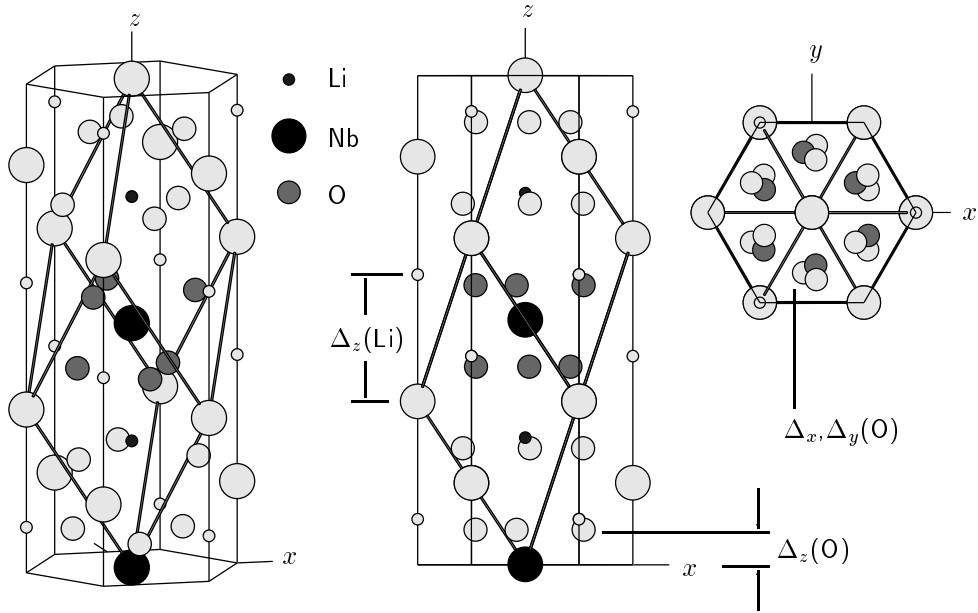,height=8.0cm}}
%\vspace{8.0cm}
\caption{
Hexagonal unit cell of ferroelectric LiNbO$_3$ (thin lines)
which includes three rhombohedral cells (thick lines).
The atoms which belong to a selected rhombohedral cell
are painted dark. Four internal crystal-structure parameters
are shown.
}
\label{fig:struct}
\end{figure}

The study of the electronic structure and lattice dynamics
from first principles was so far hindered by the relative complexity
of the structure.
In the ferroelectric phase, LiNbO$_3$ has 10 atoms in the unit cell;
the space group is $R3c$. The atomic arrangement is given
by oxygen octahedra stapled along the polar trigonal axis. Each Nb
atoms is displaced from the center of the oxygen octahedron along
the polar axis; the next octahedron (along the polar axis) is empty,
and the following one contains a Li atom, displaced from the oxygen
face along the trigonal axis. In the paraelectric phase,
the space group of LiNbO$_3$ is $R\bar{3}c$. In this
case, the Nb atoms are centered inside the oxygen octahedra, and the
Li atoms lie inside the common face of two adjacent oxygen
octahedra \cite{structure}. The primitive cell and
its internal parameters are shown in Fig.~\ref{fig:struct}.

Inbar and Cohen \cite{inbar} calculated from first principles
the total energy profile associated with the ferroelectric
instability. They have shown that a broadly spread assumption
about the instability being primarily related to the Li
displacement out of oxygen layers was not justified.
In addition to the Li displacement, the shift (with respect to Nb)
and distortion of the oxygen octahedra
play quite a crucial role and lower the total energy much more
efficiently. These results have been essentially
reproduced by Yu and Park \cite{yupark}.
In these previous studies, the ferroelectric distortion has been
simulated by an uniform scaling of the crystal structure along the linear
path between experimentally determined paraelectric and ferroelectric
structures. In the present work, we fully optimize
the structure of paraelectric and ferroelectric phases
from first principles,
adjusting the values of all internal parameters constrained only
by the crystal symmetry. In the course of that, the energetics
of individual atomic displacements can be analyzed and applied
for the zone-center lattice dynamics simulation
within the frozen-phonon scheme.
With such simulation so far missing, the attribution of different
experimentally measured zone-center phonon modes (with respect
to atomic displacement patterns) was problematic.
The extraction of normal vibration coordinates in LiNbO$_3$,
done in the present work, may be useful for the development
of reliable lattice dynamics models in this system,
including, e.g., the effects of doping.

\section*{Method}
The calculations were performed using the full potential
Linear Augmented Plane Wave method (see, e.g., \cite{singh:a})
with the addition of local orbital basis functions \cite{singh:b}
as implemented in WIEN97 FLAPW code \cite{blaha}. The exchange-correlation
was treated within the local density approximation (LDA), using the
parametrization by Perdew and Wang \cite{pw92}.
The core states were treated fully relativistically, and the semicore
and valence states were computed in a scalar relativistic approximation.
The structure optimization in the para- and ferroelectric phase
and the frozen phonon
calculations were performed using a $4\!\times\!4\!\times\!4$
special $k$-points mesh which generated 20 $k$ points
in the irreducible Brillouin zone. We tested the convergence
in the $k$-space integration
using a $6\!\times\!6\!\times\!6$ $k$-mesh (28 irreducible k-points)
and found the difference in the total energy trends, as compared
with the results on a sparser $k$-mesh, negligible for the analysis
of lattice dynamics and structure optimization.
The muffin tin radii chosen were 1.9 a.u. for Nb and 1.6 a.u. for Li and O,
close to the values used by Inbar and Cohen \cite{inbar} in their
FLAPW calculation. The convergency of the results with respect to
the number of augmented plane waves used was also controlled; we used,
on the average, 980 basis functions for each $k$-point.

\section*{Ground-state structure}

\begin{table}
\caption{Atomic positions in hexagonal coordinates}
\begin{tabular}{*{13}{c}}
\hline
phase &&
\multicolumn{3}{c}{Nb} &&
\multicolumn{3}{c}{Li} &&
\multicolumn{3}{c}{O} \\
\hline
para  && 0 & 0 & 0 && 0 & 0 & $\frac{1}{4}$ &&
$\begin{array}{c}0.049^e\\0.041^c\end{array}$ &
$\frac{1}{3}$ & $\frac{1}{12}$ \\
\hline
ferro && 0 & 0 & 0 && 0 & 0 &
$\begin{array}{c}0.283^e\\0.279^c\end{array}$&&
$\begin{array}{c}0.049^e\\0.041^c\end{array}$&
$\begin{array}{c}0.346^e\\0.344^c\end{array}$&
$\begin{array}{c}0.067^e\\0.066^c\end{array}$\\
\hline
\end{tabular}\\
$^e$exp.: Ref.~\protect\cite{structure};
$^c$calculation.
\label{tab:struc}
\end{table}

The simultaneous optimization of the volume and the $c/a$ ratio
for the paraelectric phase of LiNbO$_3$ resulted in the values
of lattice parameters (in the hexagonal setting)
$a_H$=5.1378 {\AA} and $c$=13.4987 {\AA}. As compared to the experimental
data ($a_H$=5.1483 {\AA} and $c$=13.8631 {\AA}, see Ref.~\cite{structure}),
that corresponds to a volume underestimated by $\sim3\%$ and
a $c/a$ ratio deviating by $\sim$2\% from experiment,
i.e. quite good agreement by the standards of first-principles
calculations based on the density functional theory. In the subsequent
optimization of atomic positions, we kept the lattice parameters fixed.
The fully optimized paraelectric structure was found energetically instable
with respect to the symmetry-lowering atom displacements.
The paraelectric phase has one internal coordinate
whereas the ferroelectric
phase has four. These four parameters are actually related
to the four symmetry coordinates which can be introduced
to describe the $A_1$-TO phonons. Therefore, in the course
of accumulating total energy data for our frozen phonon calculations
(see next section), we were able simultaneously to optimize
the ferroelectric ground-state structure to quite good accuracy.
Moreover, the calculated forces have been used
in the process of structure optimization.
The experimental and calculated atomic positions
(in the hexagonal coordinates, following \cite{structure})
for both paraelectric and ferroelectric phases
are given in Table~\ref{tab:struc}. The agreement between
theory and experiment in all internal parameters is quite good,
indicating a presumably nonproblematic applicability
of LDA for the study of lattice dynamics in LiNbO$_3$.
The energy difference we found between paraelectric and ferroelectric
phases is essentially the same
as determined by Inbar and Cohen \cite{inbar}.

\section*{Frozen phonons}
The $\Gamma$-TO frequencies in the ferroelectric structure
are split by symmetry into four $A_1$, five $A_2$ and
nine $E$ modes.
The frequencies of $A_1$ modes have been determined in a number
of Raman spectroscopy measurements, with a satisfactory agreement
of results \cite{kojima,ridah,schwarz}. The displacement patterns
corresponding to different modes have not yet been unambiguously
attributed, to our best knowledge. Some information to that point
has been attained, however, based on a study of the isotope effect
in Ref.~\cite{kojima}, that is addressed below.
The $A_2$ modes are both Raman and infrared silent. Our calculation
data are therefore predictive with respect to these vibrations.
The nine $E$ modes are the source of the largest controversy
in the experimental study of vibrations in LiNbO$_3$. They are
attributed differently in a number of publications. The discussion
on this controversy can be found, e.g., in Ref.~\cite{ridah}.
The first-principles description of $E$ modes remains beyond
the scope of the present study.

\begin{table}
\caption{Calculated and measured frequencies (cm$^{-1}$) of four
$A_1$-TO modes in LiNbO$_3$}
\begin{tabular}{ccccc}
\hline
&
calc. $^7$LiNbO$_3$ & exp. $^7$LiNbO$_3$ &
calc .$^6$LiNbO$_3$ & exp. $^6$LiNbO$_3$ \\
\hline
TO$_1$ & 208 & 256$^a$; 252$^b$; 251$^c$
       & 208 & 256$^a$   \\
TO$_2$ & 280 & 275$^a$; 275$^b$; 273$^c$
       & 299 & 289$^a$   \\
TO$_3$ & 344 & 332$^a$; 332$^b$; 331$^c$
       & 344 &    \\
TO$_4$ & 583 & 637$^a$; 632$^b$; 631$^c$
       & 583 & 637$^a$   \\
\hline
\end{tabular}
$^a$Ref.~\cite{kojima};
$^b$Ref.~\cite{ridah};
$^c$Ref.~\cite{schwarz};
\label{tab:freq_A1}
\end{table}

\begin{table}
\caption{Calculated eigenvectors of four $A_1$ modes}
\begin{tabular}{crr@{.}lr@{.}lr@{.}l
                  r@{.}lr@{.}lr@{.}lr@{.}lr@{.}l}
\hline
Mode &
 & \multicolumn{2}{c}{Nb(2$\times$)}
 & \multicolumn{2}{c}{Li(2$\times$)}
 & \multicolumn{2}{c}{O}  & \multicolumn{2}{c}{O}
 & \multicolumn{2}{c}{O}  & \multicolumn{2}{c}{O}
 & \multicolumn{2}{c}{O}  & \multicolumn{2}{c}{O} \\
\hline
    &
$x$ & \multicolumn{2}{c}{0} &
      \multicolumn{2}{c}{0} &
         0&07 & $-$0&04 & $-$0&03 &    0&07 & $-$0&03 & $-$0&04 \\*[-2mm]
TO$_1$ &
$y$ & \multicolumn{2}{c}{0} &
      \multicolumn{2}{c}{0} &
         0&01 &    0&06 & $-$0&07 & $-$0&01 &    0&07 & $-$0&06 \\*[-2mm]
    &
$z$ &  ~~0&39 &        0&09 &
      $-$0&33 & $-$0&33 & $-$0&33 & $-$0&33 & $-$0&33 & $-$0&33 \\
\hline
    &
$x$ & \multicolumn{2}{c}{0} &
      \multicolumn{2}{c}{0} &
      $-$0&01 &    0&01 &    0&00 & $-$0&01 &    0&00 &    0&01 \\*[-2mm]
TO$_2$ &
$y$ & \multicolumn{2}{c}{0} &
      \multicolumn{2}{c}{0} &
      $-$0&01 & $-$0&01 &    0&01 &    0&01 & $-$0&01 &    0&01 \\*[-2mm]
    &
$z$ &    0&18 &     $-$0&68 &
         0&00 &    0&00 &    0&00 &    0&00 &    0&00 &    0&00 \\
\hline
    &
$x$ & \multicolumn{2}{c}{0} &
      \multicolumn{2}{c}{0} &
      $-$0&14 & $-$0&27 &    0&40 & $-$0&14 &    0&40 & $-$0&27 \\*[-2mm]
TO$_3$ &
$y$ & \multicolumn{2}{c}{0} &
      \multicolumn{2}{c}{0} &
         0&38 & $-$0&31 & $-$0&08 & $-$0&38 &    0&08 &    0&31 \\*[-2mm]
    &
$z$ &    0&02 &        0&00 &
      $-$0&01 & $-$0&01 & $-$0&01 & $-$0&01 & $-$0&01 & $-$0&01 \\
\hline
    &
$x$ & \multicolumn{2}{c}{0} &
      \multicolumn{2}{c}{0} &
      $-$0&38 &    0&31 &    0&07 & $-$0&38 &   0&07 &    0&31 \\*[-2mm]
TO$_4$ &
$y$ & \multicolumn{2}{c}{0} &
      \multicolumn{2}{c}{0} &
      $-$0&14 & $-$0&26 &    0&39 &    0&14 & $-$0&39 &    0&26 \\*[-2mm]
    &
$z$ &    0&06 &        0&04 &
      $-$0&06 & $-$0&06 & $-$0&06 & $-$0&06 & $-$0&06 & $-$0&06 \\
\hline
\end{tabular}
\label{tab:evec_A1}
\end{table}

The calculation of the force constants to be used in the soft-phonon
calculation typically involves a second-order fit over a number
of data points, corresponding to different displacements within
a given symmetry constraint. For $A_1$ modes, we had
to consider more than ninety different geometries in order to obtain
a satisfactory total-energy fit in a sense that the results
remain relatively unaffected by the addition of extra total energy data.
The latter applies at least to two (TO$_2$ and TO$_3$) of the four modes.
The calculated frequencies of the $A_1$ modes are shown
in Table~\ref{tab:freq_A1} in comparison with the experimental data.
The agreement is very good for TO$_2$ and TO$_3$ modes.
Taken together with the above mentioned stability of calculated
frequencies with respect to improving the total-energy fit,
this indicates that these two vibration modes are with high accuracy
harmonic. The corresponding eigenvectors are shown
in Table~\ref{tab:evec_A1}. (Note that the displacement of both Nb
atoms and both Li atoms in the unit cell is identical
within the $A_1$ modes.) One can see
that TO$_2$ is essentially the $z$-vibration of Li ions
with respect to a (relatively rigid) rest of a crystal.
Actually this is the only $A_1$ mode with a substantial amount
of Li movement. It is clearly seen in the experimentally measured
frequencies of $^6$Li-doped LiNbO$_3$ \cite{kojima} that only the
TO$_2$ mode exhibits an isotope effect, increasing its frequency
by 14 cm$^{-1}$ (see Table~\ref{tab:freq_A1}). Our calculation
of vibration frequencies with the decreased mass of Li ion
as well indicates that only the TO$_2$ mode is affected,
its frequency being increased by 19 cm$^{-1}$.

The displacement pattern in the TO$_1$ mode has some resemblance
to that in the soft mode of cubic perovskites, like e.g. KNbO$_3$:
essentially, Nb vibrates in antiphase with the oxygen sublattice
along the trigonal axis, leaving Li relatively static.
As follows from the frozen-phonon treatment in cubic KNbO$_3$
(see, e.g., Ref.~\cite{sb92,phonon}), this mode is instable against
off-center displacements and hence exhibits in the harmonic
approximation an imaginary frequency. But even when stabilized
by an appropriate symmetry lowering, as was calculated
for example for tetragonal \cite{phonon}
or orthorhombic \cite{ortho} KNbO$_3$, the mode
in question roughly maintains its original displacement pattern.
In LiNbO$_3$, the TO$_1$ is the ultimately stabilized
soft mode of the paraelectric phase. We analyzed the total energy
as function of atomic displacements
consistent with the TO$_1$ eigenvector and found noticeable
deviations from the parabolic behaviour. Because of this,
the calculated harmonic frequency strongly differs from
the experimental numbers.

\begin{figure}
\centerline{\epsfig{file=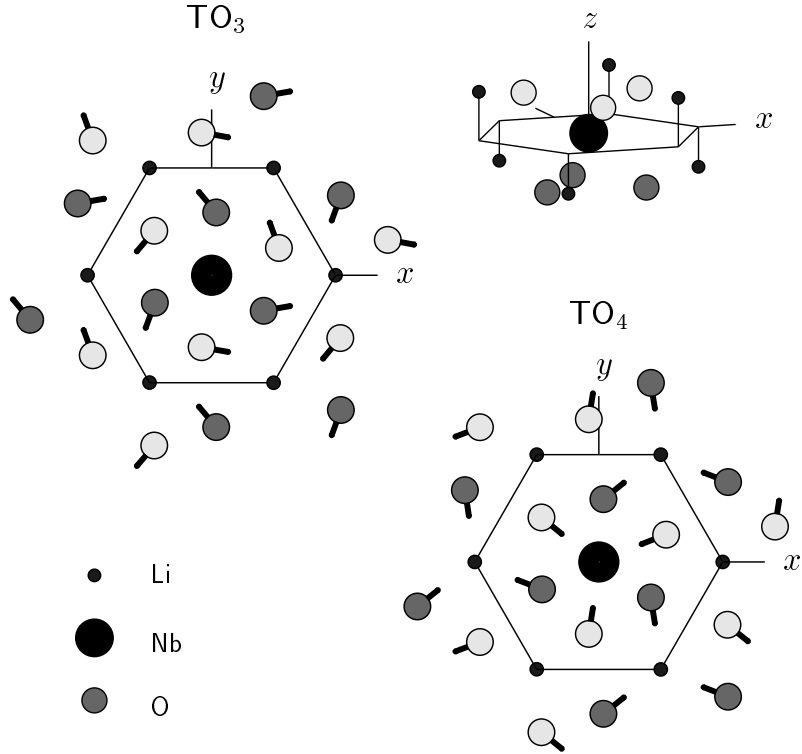,height=10.0cm}}
%\vspace{10.0cm}
\caption{
Top view of the displacement patterns in the $A_1$ TO$_3$ and
TO$_4$ phonon modes. Only the atoms closest to the base
of the hexagonal unit cell (as in Fig.~\ref{fig:struct})
are shown. Oxygen atoms above the base are painted light grey
and below the plane dark grey (see inset on the top).
}
\label{fig:phonon}
\end{figure}

Another example of a strongly anharmonic mode is the TO$_4$ mode.
Along with TO$_3$, it is visualized in Fig.~\ref{fig:phonon}
in a top view (along the $z$-axis). The $z$-displacements
are negligible in these two modes, and practically only
oxygen ions are participating in them. The difference between
these modes is the following: in TO$_3$, the whole oxygen octahedra
are tilted as essentially rigid objects, that is a relatively
soft and harmonic vibration. In the TO$_4$ mode, the torsion
of individual octahedra takes place, that costs much higher
energy and has a strong anharmonic contribution.

\begin{table}
\caption{Calculated frequencies and eigenvectors of five $A_2$ modes}
\begin{tabular}{crr@{.}lr@{.}l@{\hspace*{-8mm}}
                r@{.}lr@{.}lr@{.}lr@{.}lr@{.}lr@{.}l}
\hline
$\omega$ (cm$^{-1}$) &
 & \multicolumn{2}{c}{\hspace*{-1mm}Nb($+/-$)\hspace*{-1mm}}
 & \multicolumn{2}{c}{\hspace*{-1mm}Li($+/-$)\hspace*{-1mm}}
 & \multicolumn{2}{c}{O}  & \multicolumn{2}{c}{O}
 & \multicolumn{2}{c}{O}  & \multicolumn{2}{c}{O}
 & \multicolumn{2}{c}{O}  & \multicolumn{2}{c}{O} \\
\hline
    &
$x$ & \multicolumn{2}{c}{0} &
      \multicolumn{2}{c}{0} &
      $-$0&03 &    0&02 &    0&01 &    0&03 & $-$0&01 & $-$0&02 \\*[-2mm]
153 &
$y$ & \multicolumn{2}{c}{0} &
      \multicolumn{2}{c}{0} &
         0&00 & $-$0&03 &    0&03 &    0&00 &    0&03 & $-$0&03 \\*[-2mm]
    &
$z$ &    0&24 & $-$0&66 &
         0&05 &    0&05 &    0&05 & $-$0&05 & $-$0&05 & $-$0&05 \\
\hline
    &
$x$ & \multicolumn{2}{c}{0} &
      \multicolumn{2}{c}{0} &
      $-$0&08 &    0&24 & $-$0&16 &    0&08 &    0&16 & $-$0&24 \\*[-2mm]
287 &
$y$ & \multicolumn{2}{c}{0} &
      \multicolumn{2}{c}{0} &
      $-$0&23 &    0&05 &    0&18 & $-$0&23 &    0&18 &    0&05 \\*[-2mm]
    &
$z$ & $-$0&51 & $-$0&14 &
         0&12 &    0&12 &    0&12 & $-$0&12 & $-$0&12 & $-$0&12 \\
\hline
    &
$x$ & \multicolumn{2}{c}{0} &
      \multicolumn{2}{c}{0} &
      $-$0&23 & $-$0&04 &    0&26 &    0&23 & $-$0&26 &    0&04 \\*[-2mm]
417 &
$y$ & \multicolumn{2}{c}{0} &
      \multicolumn{2}{c}{0} &
         0&17 & $-$0&28 &    0&11 &    0&17 &    0&11 & $-$0&28 \\*[-2mm]
    &
$z$ &    0&03 &    0&11 &
         0&28 &    0&28 &    0&28 & $-$0&28 & $-$0&28 & $-$0&28 \\
\hline
    &
$x$ & \multicolumn{2}{c}{0} &
      \multicolumn{2}{c}{0} &
         0&01 & $-$0&25 &    0&24 & $-$0&01 & $-$0&24 &    0&25 \\*[-2mm]
440 &
$y$ & \multicolumn{2}{c}{0} &
      \multicolumn{2}{c}{0} &
         0&28 & $-$0&13 & $-$0&15 &    0&28 & $-$0&15 & $-$0&13 \\*[-2mm]
    &
$z$ & $-$0&42 & $-$0&19 &
      $-$0&12 & $-$0&12 & $-$0&12 &    0&12 &    0&12 &    0&12 \\
\hline
    &
$x$ & \multicolumn{2}{c}{0} &
      \multicolumn{2}{c}{0} &
      $-$0&33 &    0&21 &    0&12 &    0&33 & $-$0&12 & $-$0&21 \\*[-2mm]
883&
$y$ & \multicolumn{2}{c}{0} &
      \multicolumn{2}{c}{0} &
      $-$0&06 & $-$0&26 &    0&31 & $-$0&06 &    0&31 & $-$0&26 \\*[-2mm]
    &
$z$ &    0&06 &    0&02 &
      $-$0&23 & $-$0&23 & $-$0&23 &    0&23 &    0&23 &    0&23 \\
\hline
\end{tabular}
\label{tab:evec_A2}
\end{table}

For the $A_2$ modes, no experimental information is available
by means of Raman nor infrared spectroscopy, so our results
are actually a theory prediction. We used 127 different
geometries to obtain an accurate second-order total energy fit
in the 5-dimensional space of symmetry coordinates.
Calculated frequencies and
eigenvectors are shown in Table~\ref{tab:evec_A2}.
Note that in contrast
to $A_1$ modes, both Nb and both Li atoms are now moving
in antiphase. Apart from the softest mode which includes
essentially the Nb vs. Li antiphase $z$-movement, and the hardest one,
which is again a distortion of oxygen octahedra (but
now including the $z$-stretching as well), three intermediate
modes are have contributions from $z$-
as well as $xy$-displacements of all three atomic constituents.

Summarizing, we performed from first principles in a LDA-based
calculation the optimization of the ground structure of
ferroelectric LiNbO$_3$ and calculated the frequencies and
eigenvectors of $A_1$ and $A_2$ $\Gamma$-TO modes.
Large anharmonic contributions were found
for two of four $A_1$ modes. These results may present a basis
for a subsequent treatment of lattice dynamics models in
related systems and/or of anharmonic effects. The study
of $E$ modes is now in progress.

\section*{Acknowledgements}
The work was supported by the German research Society
(SFB 225; graduate college). A.~P. is grateful to
R.~Cohen for useful discussions. The authors appreciate
the usefulness of the crystal structure visualization
software written and provided by M.~Methfessel.

\end{document}